\documentclass[twocolumn]{aastex631}

\usepackage{amsmath}
\usepackage{booktabs}
\begin{document}

\title{A new lever on exoplanetary B fields: measuring heavy ion velocities}

\correspondingauthor{Arjun B. Savel}
\email{asavel@umd.edu}

\newcommand\umd{\affiliation{Astronomy Department, University of Maryland, College Park, 4296 Stadium Dr., College Park, MD 207842 USA}}

\newcommand\ucr{\affiliation{Department of Earth Sciences, University of California, Riverside, Riverside, CA,
USA.}}

\author[0000-0002-2454-768X]{Arjun B. Savel}
\umd

\author[0000-0002-6980-052X]{Hayley Beltz}
\umd

\author[0000-0002-9258-5311]{Thaddeus D. Komacek}
\umd

\author[0000-0002-8163-4608]{Shang-Min Tsai}
\ucr

\author[0000-0002-1337-9051]{Eliza M.-R. Kempton}
\umd



\begin{abstract}
Magnetic fields connect an array of planetary processes, from atmospheric escape to interior convection. Despite their importance, exoplanet magnetic fields are largely unconstrained by both theory and observation. In this Letter, we propose a novel method for constraining the B field strength of hot gas giants: comparing the velocities of heavy ions and neutral gas with high-resolution spectroscopy. The core concept of this method is that ions are directly deflected by magnetic fields. While neutrals are also affected by B fields via friction with field-accelerated ions, ionic gas should be more strongly coupled to the underlying magnetic field than bulk neutral flow. Hence, measuring the difference between the two velocities yields rough constraints on the B field, provided an estimate of the stellar UV flux is known. We demonstrate that heavy ions are particularly well suited for this technique, because they are less likely to be entrained in complex hydrodynamic outflows than their lighter counterparts. We perform a proof-of-concept calculation with Ba II, an ion whose velocity has been repeatedly measured at high confidence with high-resolution spectroscopy. Our work shows that a 10G magnetic field would produce $\sim$ km\,s$^{-1}$ ion--neutral velocity differences at a microbar, whereas a 50G magnetic field would produce $\sim$ 20~km\,s$^{-1}$ velocity difference. With new leverage on magnetic fields, we will be able to investigate magnetic field generation in the extreme edge cases of hot gas giants, with wide-ranging consequences for planetary interior structure, dynamo theory, and habitability.
\end{abstract}



\section{Introduction} \label{sec:intro}
For the foreseeable future, exoplanets can only be investigated by remote sensing. Of the exoplanetary properties accessible by remote sensing, magnetic fields are one of the most fundamental yet least known. While they influence many aspects of planetary science---such as heat redistribution \citep[e.g.,][]{menou2012magnetic}, habitability \citep[e.g.,][]{meadows2018factors}, evolution \citep[e.g.,][]{driscoll2015tidal}, and interior structure \citep[e.g.,][]{zarka2015magnetospheric}---we simply do not know their strengths \citep[e.g.,][]{brain2024exoplanet}.

Theory does not provide much guidance on B field strengths for hot gas giants. Depending on the dynamo scaling law chosen, these planets' B field strengths could be on the order of Jupiter's \citep[$\sim$ 4~G at the equator; ][]{sanchez2004magnetic,stevens2005magnetospheric,zaghoo2018size}, an order of magnitude smaller \citep[e.g.,][]{christensen2010dynamo}, or nearly two orders of magnitude larger \citep{yadav2017estimating}. 

\begin{table*}
\centering

\caption{Ba II detections in exoplanet atmospheres}\label{table:barium_detections}
\begin{tabular}{lccccc}
\toprule
Reference & Planet & Ba II S/N & $u_{\rm Ba\,II}$ (km/s) & Fe I S/N &  $u_{\rm Fe\,I}$ (km/s) \\
\midrule
\cite{silva2022detection}\footnote{For this study, we only show results from the second of two nights of observation.} & WASP-76b & 7.4 & $-4.00\pm0.58$ & 22.2 & $-3.92\pm0.34$ \\
\cite{silva2022detection} & WASP-121b & 10.6 & $-1.9\pm0.5$ & 16.3 & $-2.6\pm0.3$ \\
\cite{borsato2023mantis}\footnote{This study technically reports t-test significance, as opposed to signal-to-noise ratios.} & KELT-9b & 4.5 & $-19.0$ & 15.3 & $-19.0$ \\
\cite{pelletier2023vanadium}\footnote{We derive the velocities and errors by fitting a Gaussian function to each species' cross-correlation function (provided by Pelletier, priv. comm.) in the planetary rest frame.} & WASP-76b & 4.95 & $-8.80\pm0.59$ & 12.94 & $-6.39\pm0.29$\\
\cite{prinoth2023time} & WASP-189b & 9.12 & $-0.97\pm1.14$ & 22.7 & $-4.31\pm0.52$  \\
\bottomrule
\end{tabular}

\end{table*}

Observations have \textit{also} not provided strong magnetic field constraints. Extensive efforts  have been made to connect planetary magnetic field strength to, e.g., the circulation of neutral gas \citep{perna2010magnetic,menou2012magnetic,rauscher2012general,batygin2013magnetically,rauscher2013three,rogers2014magnetohydrodynamic,rogers2014magnetic,rogers2017hottest,beltz2021exploring,knierim2022shallowness} and their velocities measured with high-resolution spectroscopy \citep{kempton2012constraining,beltz2022magnetic, beltz2023magnetic}, UV transit timing asymmetries from bow shocks \citep[e.g.,][]{vidotto2011prospects, kislyakova2014magnetic}, radio auroral emission from the planet \citep[e.g.,][]{zarka2007plasma,narang2024ugmrt,shiohira2024search} and star \citep[e.g.,][]{pineda2023coherent}, and helium line velocity measurements \citep{schreyer2024using} and spectropolarimetry \citep{oklopvcic2020detecting}. Despite these efforts, success has been isolated \citep{ben2022signatures}, indirect \citep{cauley2019magnetic}, or difficult to reconcile across repeated observations \citep[e.g.,][]{lally2022reassessing,turner2023follow}.

Between theory and observations, B field strengths are unknown to orders of magnitude uncertainty. There is clear need for novel techniques to provide more leverage on this problem. In this Letter, we present a proof of concept: using measured heavy ion--neutral velocity differences to infer exoplanet magnetic field strengths. Unlike neutrals, ions directly ``feel'' the magnetic field. And unlike light ions, heavy ions require more energy input to be entrained in an escaping flow, making them more robust tracers of field lines. This technique is promising by virtue of its observational connections: high-resolution ground-based spectroscopy \citep{snellen2010orbital,birkby2018exoplanet} has yielded numerous heavy ion detections and velocity measurements in just the past few years \citep[Table~\ref{table:barium_detections};][]{silva2022detection,borsato2023mantis,pelletier2023vanadium, prinoth2023time}.

We present our basic magnetohydrodynamic (MHD) arguments in Section~\ref{sec:mhd}. We then briefly demonstrate why heavy ions are less likely to be lofted into an atmosphere outflow (Section~\ref{sec:escape}). Section~\ref{sec:proof} lays out our primary proof of concept, calculating the ion--neutral velocity difference for a representative ultra-hot Jupiter atmosphere. Finally, we summarize and provide thoughts on further work in Section~\ref{sec:conclusion}.

\section{Magnetohydrodynamics}\label{sec:mhd}
Below, we set up a physical system similar to MHD approaches that calculate ``magnetic drag.'' Such models consider the drag that neutral particles experience in a partially ionized atmosphere as they collide with charged particles, the latter of which experience Lorentz forces from field lines \citep[e.g.,][]{zhu2005self,perna2010magnetic,rauscher2013three,beltz2021exploring}. We here focus not on the motion of neutrals as arrested by ions, but rather the relative motions of the \textit{ions themselves}.\footnote{We emphasize that an ion--neutral force difference is the mechanism that produces magnetic drag. Therefore, a decoupling of the neutral and charged gas velocity structures is consistent with previous exoplanet MHD studies. What remains to be seen is how tightly these flow structures are coupled.} If the flow structures are not tightly coupled, then strong ion--neutral velocity differences, or ``slip velocities'' \citep[e.g.,][]{hopkins24micro}, would correspond to strong planetary magnetic fields.

By combining the electron and ion momentum equations, \cite{koskinen2014electrodynamics} find 

\begin{equation}
    (\boldsymbol{u}_p - \boldsymbol{u}_n) \approx \frac{\boldsymbol{j} \times \boldsymbol{B}}{n_pm_i\nu_{in}}
\end{equation}
for plasma velocity $\boldsymbol{u}_p$, neutral velocity $\boldsymbol{u}_n$, current $\boldsymbol{j}$, magnetic field $\boldsymbol{B}$, plasma number density $n_p$, ion mass $m_i$, and collision frequency $\nu_{in}$.

Evaluating $\boldsymbol{j}$ is non-trivial. We can make the simplifying assumption of ``resistive MHD,'' equivalent to assuming that the conductivity tensor can be expressed as a scalar. Then the current is \citep[e.g.,][]{perna2010magnetic, koskinen2014electrodynamics} 

\begin{equation}
    j = \sigma u_n B,
\end{equation}
for conductivity $\sigma$, considering only the component of the current that is entirely orthogonal to the (assumed deep-seated, dipolar, aligned with the rotation of the planet) B field. Therefore, the difference in velocities between ions and neutrals can be written as 

\begin{equation}
    |\boldsymbol{u}_p - \boldsymbol{u}_n| \approx \frac{\sigma u_n B^2}{n_pm_i\nu_{in}}.
\end{equation}

We now calculate the ion--neutral collision frequency as (analagously to, e.g., \citealt{helling2021understanding})

\begin{equation}
    \nu_{in} = \sigma_{\rm coll} n_{\rm gas} v_{\rm rel},
\end{equation}
where $\sigma_{\rm coll}$ is the cross-section for ion--neutral collisions, $n_{\rm gas}$ is the gas density, and $v_{\rm rel}$ is the relative velocity between the neutrals and ions. 

In the high-temperature limit, the thermal velocity is expected to be greater than the bulk flow difference $u_p - u_n$.\footnote{This approximation breaks down when the slip speed exceeds the thermal speed, at B fields of roughly 100~G. For superthermal slip speeds, bulk flow will contribute to a higher collision rate, and the system may be driven toward \textit{subthermal} slip velocities \citep{hillier2024ambipolar,hopkins24micro}.} We hence consider our relative velocity to be the thermal velocity of hydrogen, which is greater than the heavy ions we consider in a given collision:\footnote{We remark for this calculation, as does \cite{helling2021understanding}, that hot enough atmospheres are not composed of pure molecular hydrogen---an appreciable fraction of $\rm H_2$ thermally dissociates into atomic H \citep{bell2018increased,komacek2018effects,tan2019atmospheric,mansfield2020evidence,roth2021pseudo}. The primary impact of dissociation would be to decrease the mass of the dominant neutral particle, thereby increasing the thermal velocity, increasing the collision frequency, and therefore decreasing the slip velocity for a given B field.}

\begin{equation}
     v_{\rm rel} \approx \sqrt{\frac{k_BT}{m_{H_2}}}.
\end{equation}

We also assume that a heavy ion is much larger than a neutral atom in an average collision:

\begin{equation}
    \sigma_{\rm coll} \approx \pi r_i^2.
\end{equation}

Taken together, the slip velocity can therefore be written as 

\begin{equation}
    |\boldsymbol{u}_p - \boldsymbol{u}_n| \approx \frac{\sigma u_n B^2}{n_pm_in_{\rm gas}\pi r_i^2}\sqrt{\frac{m_{H_2}}{k_BT}}.
\end{equation}

We also note that, from \cite{perna2010magnetic}, the conductivity can be written simply as 

\begin{equation}
    \sigma = \frac{n_e e^2}{m_en_n \mathcal{M}_{en}},
\end{equation}
where $n_e$ is the electron density, $e$ is the electric charge, $m_e$ is the electron mass, $n_n$ is the neutral gas number density, and $\mathcal{M}_{en}$ is the electron--neutral momentum transfer coupling term (in units of speed). \cite{draine1983magnetohydrodynamic} provide the following estimate for $\mathcal{M}_{en}$:

\begin{equation}
    \mathcal{M}_{en} = 10^{-15}\sqrt{\frac{128k_BT}{9\pi m_e}}.
\end{equation}
Hence

\begin{equation}
    \sigma = 10^{15}\frac{n_e e^2}{m_en_n}\sqrt{\frac{9\pi m_e}{128k_BT}}.
\end{equation}

The full expression for the ion--neutral velocity difference is slightly unwieldy, but it is worth examining. 

Altogether, noting that $n_e = n_p$ in the \cite{koskinen2014electrodynamics} formalism, we have:

\begin{equation}\label{eq:central_equation}
    |\boldsymbol{u}_p - \boldsymbol{u}_n| \approx 10^{15}\frac{u_n e^2 B^2}{m_in_nn_{\rm gas}r_i^2 k_BT}\sqrt{\frac{9m_{H_2}}{128\pi m_e}}.
\end{equation}

We now identify a few noteworthy scalings. The larger and more massive the ion, the stronger the coupling is to the neutral flow. Hence, some of the heaviest ions detected in exoplanet atmospheres to date (such as Tb II; \citealt{borsato2023mantis}) may not be ideal candidates for tracing B fields. We argue, however, that the lightest ions are not preferable for tracing B fields because they may be entrained in hydrodynamic outflows (explored further in Section~\ref{sec:escape}), so a moderately massive ion is ideal.

We also highlight the strong dependence on gas density in Equation~\ref{eq:central_equation}. The lower the density, the less strongly the ion and neutral flows are coupled. This scaling is sensible, because lower density leads to fewer collisions. In hydrostatic atmospheres, gas density falls off exponentially with altitude. So, the best candidate ions for trace B fields have strong opacity, causing the gas to become optically thick at low pressures, in turn allowing spectroscopy to probe low-pressure regions. The tradeoff is that if too low pressures are probed, one may be observing the upper atmosphere and therefore the non-hydrostatic outflow (which begins, to order of magnitude, at 1 nanobar for a standard hot Jupiter and scales linearly with surface gravity; \citealt{murray2009atmospheric}).

Also note the dependence on the neutral gas number density, $n_n$. The higher this quantity is (i.e., the lower the electron number density, $n_e$), the weaker the relationship between B field and ion--neutral velocity difference. Therefore, a more ionized atmosphere will better enable ions to trace B fields.

Finally, note the strong dependence on $B$. Here is where the strength of this proposed technique lies: regardless of assumptions about hydrogen dissociation, or the exact free electron population, the ion--neutral velocity difference is highly dependent on the magnetic field. So, while precisely constraining B fields will require more theoretical work, a high velocity difference can reasonably imply a strong B field, and a low velocity difference can imply a weak B field.

For very strong B fields, this slip velocity can become very large. For such high velocities, we require an additional Lorentz factor:

\begin{equation}\label{eq:relativistic version}
    |\gamma_p\boldsymbol{u}_p - \boldsymbol{u}_n| \approx 10^{15}\frac{u_n e^2 B^2}{m_in_nn_{\rm gas}r_i^2 k_BT}\sqrt{\frac{9m_{H_2}}{128\pi m_e}},
\end{equation}
where

\begin{equation}
    \gamma_p = \frac{1}{\sqrt{1 - \frac{u_p^2}{c^2}}}.
\end{equation}

Equation~\ref{eq:relativistic version} implies that, for atmospheres with very high B fields and ionization fractions, the (relativistic) kinetic energy of the ions in the atmosphere may exceed the binding energy of a given atmospheric shell. This effect would of course violate hydrostatic equilibrium, resulting in strong vertical velocities. Strong vertical velocities have been inferred via resolved alkali lines at high resolution \citep[e.g.,][]{seidel2020wind}, though such signatures are likely connected to hydrodynamic outflow driven by instellation. If ion speeds are not regulated by, e.g., current instabilities \citep[e.g.,][]{norman1978kinetic, hopkins24micro}, relativistic plasma velocity would also be associated with inverse-Compton scattering and synchrotron radiation from electrons rapidly accelerating along (assumed curved) B field lines. Presently, there is no evidence of such radiation.

\section{Light ions are not preferable probes of B fields}\label{sec:escape}

 As presented, the results of Section~\ref{sec:mhd} hold for any ion. However, not all ions would be \textit{effective} B field probes---an ideal one remains in the hydrostatic interior. Even if ions roughly follow field lines in the upper atmosphere \citep{owen2019atmospheric}, being exposed to the stellar wind introduces a host of other complicating ion processes \citep[e.g.,][]{gronoff2020atmospheric}. Furthermore, ion velocity structures would be easier to interpret without having to consider a hydrodynamic outflow \citep[e.g.,][]{yelle2004aeronomy, murray2009atmospheric}. We maintain, therefore, that it is simpler to interpret ions following field lines if the ions are not entrained in these outflows.

   \begin{figure*}
     \centering
     \includegraphics[scale=0.69]{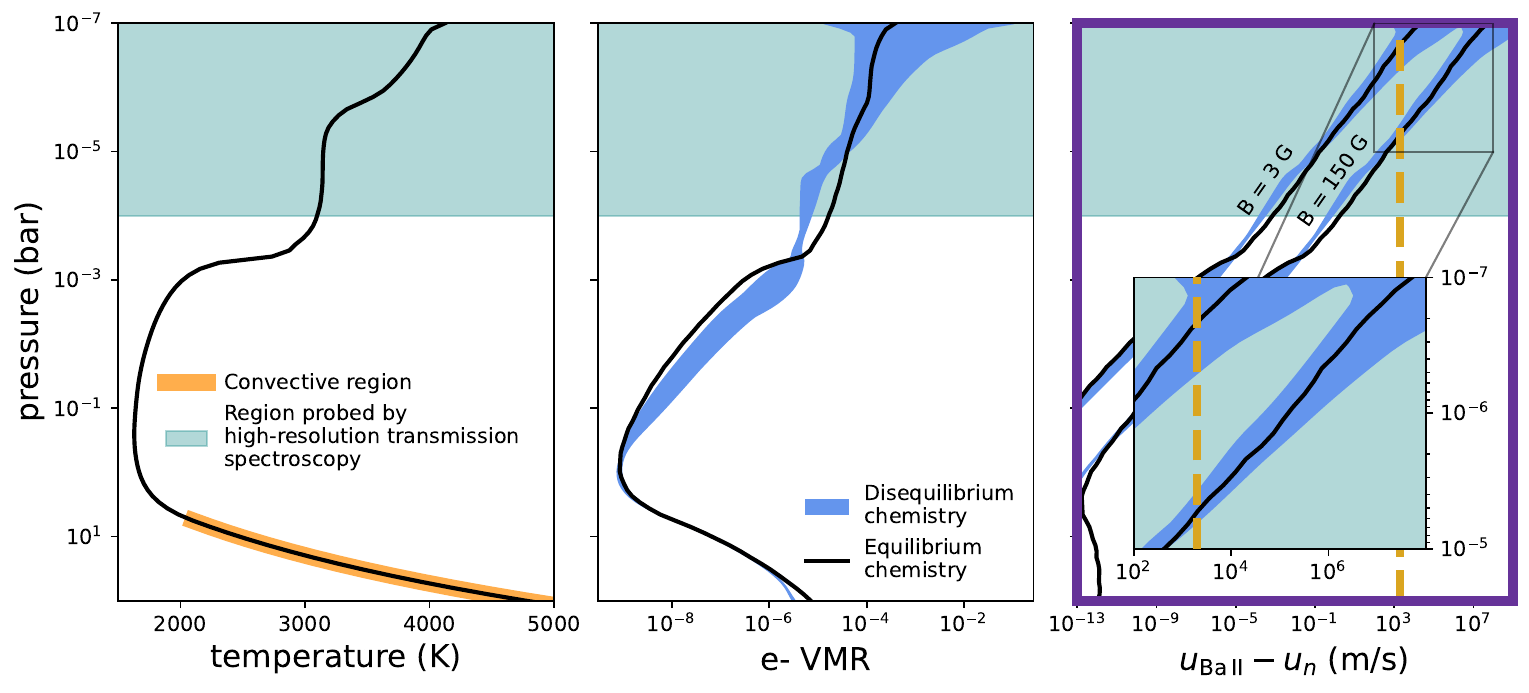}
     \caption{Calculated properties of the model atmosphere described in Table ~\ref{table:parameters}. Left: the limb temperature--pressure structure calculated with \texttt{HELIOS} \citep{malik2017helios, malik2019self}. Regions where the atmosphere is convective are highlighted in orange. Pressures generally probed in high-resolution transmission spectroscopy \citep[e.g.,][]{kempton2014high,gandhi2020seeing,hood2020prospects, wardenier2023modelling} are filled in green. Middle: free electron volume mixing ratio calculated by \texttt{FastChem} equilibrium \citep[in black;][]{kitzmann2024fastchem} and \texttt{VULCAN} disequilibrium \citep[shaded blue;][]{tsai2021comparative}  chemistry. We run \texttt{VULCAN} over a wide range of parameters---hence, there is a width to this region. Right: ion--neutral velocity difference (Equation~\ref{eq:central_equation}), calculated under equilibrium and disequilibrium chemistry assuming two B field strengths. The gold line indicates the approximate minimum detectable velocity difference via high-resolution spectroscopy \citep[e.g.,][]{ehrenreich2020nightside}. The width of the disequilibrium chemistry grid is due almost entirely to the changing input stellar flux.}
     \label{fig:helios_models}
 \end{figure*}

\begin{table}[t]
    \centering
    \caption{Parameters of the 1D ultra-hot Jupiter model presented in Section~\ref{sec:proof}}
    \label{table:parameters}
    \begin{tabular}{lr}
        \toprule
        Parameter & Value \\
        \midrule
        $g$ (cm\,s$^{-2}$) & 900\\
        $a$ (AU) & 0.025\\
        $R$ ($R_{\text{Jup}}$) & 1.8 \\
        $T_{\text{int}}$ (K)\footnote{Intrinsic temperature, calculated with the formalism of \cite{thorngren2019intrinsic}.} & 560\\
        $T_{\text{star}}$ (K) & 6500\\
        $R_{\text{star}}$ ($R_{\odot}$) & 1.5\\
        $K_{\rm zz}$ (cm$^{2}$\,s$^{-1}$) & [$10^8$, $10^9$, $10^{10}$, $10^{11}$, $10^{12}$, $10^{13}$]\\
         Stellar XUV spectrum & F5V star,\footnote{Drawn from \cite{rugheimer2013spectral}.} HD 189733\footnote{Drawn from \cite{bourrier2020moves}.}\\
        \bottomrule
    \end{tabular}
\end{table}

 We argue that the \textit{heavier} an ion is, the more likely it is to not be entrained in outflows. Atmospheric escape is of course a very complex process, with 3D numerical simulations revealing detailed geometric structure \citep[e.g.,][]{bourrier20133d,tripathi2015simulated,debrecht2019photoevaporative} and the stellar environment ultimately controlling many observables \citep[e.g.,][]{owen2023fundamentals}. We present below a simple motivating argument in the planet frame.

 To briefly demonstrate why a heavy ion would be less likely to participate in an atmospheric outflow, we consider the case of XUV (1–120 nm)-driven hydrodynamic (i.e., photoevaporative) escape. In this regime, escaping hydrogen can drag heavier elements along \citep{hunten1987mass}. \cite{schaefer2016predictions} define the ``critical XUV flux'' required for this hydrogen drag to occur as 

 \begin{equation}\label{eq:critical_flux}
     F_{\rm XUV}^{\rm crit} = \frac{4b_{12}\mathcal{V}^2_1}{\epsilon k_B T R_{\rm P}}((\mu_2/\mu_1) - 1)X_1,
 \end{equation}
where $b_{12}$ is the binary diffusion coefficient \citep{mason1970diffusion,zahnle1986mass} between the primary escaping species (1) and the dragged species (2), $\mathcal{V}$ is the potential energy of a single particle of the primary escaping species, $\epsilon$ is an efficiency factor that is usually between 0.15 and 0.3 \citep{watson1981dynamics,kasting1983loss,chassefiere1996hydrodynamic,tian2009thermal, schaefer2016predictions}, $R_{\rm P}$ is the planetary radius, $\mu_i$ is the mean molecular weight of species $i$, and $X_1$ is the molar concentration of the primary escaping species. 
 
 As an example, we take the ratio of the critical fluxes required to drag Ba II and Ca II. Assuming that the two have similar binary diffusion coefficients with respect to the primary escaping species,

 \begin{equation}
     \frac{F_{\rm XUV, Ba \,II}^{\rm crit}}{F_{\rm XUV, Ca \,II}^{\rm crit}} = \frac{(\mu_{\rm Ba \,II}/\mu_1) - 1}{(\mu_{\rm Ca\, II}/\mu_1) - 1}.
 \end{equation}

If these ions are dragged by atomic H, this ratio evaluates to $\sim 3.5$. That is, a factor of $\sim 3.5$ higher XUV flux is required for Ba II escape vs. Ca II escape. This factor is marginally greater than the factor of $\sim 3$ in both the variation of XUV flux across intermediate-mass main-sequence stars \citep[e.g.,][]{fossati2018extreme} and the precision attainable with XUV reconstructions from observations \citep[e.g.,][]{schaefer2016predictions, youngblood2016muscles}. Notably, Ca II has been observed with absorption so strong as to be detected by fitting individual line profiles (as opposed to via cross-correlation), with transit depths on the scale of the planetary Roche lobe \citep[e.g.,][]{deibert2021detection,tabernero2021espresso, silva2022detection}, whereas Ba II has not. Because the transit depths in the Ba II line cores are smaller than the planetary Roche lobe, the atmospheric layers that Ba II spectral lines probe appear to be within the hydrostatic atmosphere, whereas Ca II line cores almost certainly probe outflowing regions.

The previous conclusion hinges on a mean molecular weight argument. Any other heavy ion will fare similarly---it will remain in the hydrostatic interior, where it will more straightforwardly follow field lines, under a broader range of conditions than a light ion.

\section{Proof of concept: Barium ion in an ultra-hot Jupiter}\label{sec:proof}
Sections~\ref{sec:mhd} and ~\ref{sec:escape} motivate using a moderately heavy, high-opacity, low-abundance ion to probe exoplanetary B fields. We now flesh out our argument with a case study: Ba II in an ultra-hot Jupiter atmosphere. These atmospheres are hot enough to be appreciably partially (thermally) ionized \citep[e.g.,][]{arcangeli2018h,parmentier2018thermal}, making them excellent test cases for our ion-focused framework.

%If we plug in Barium+ at T=3000~K and P=1 microbar and 3 Gauss, and a background flow of 5 km/s, we get a neutral--ion evelocity difference of roughly 9 m/s. This is not measurable with current instruments. But the B field dependence saves us: 50 Gauss B field puts us at a 2.6 km/s velocity difference.

We begin by calculating the limb temperature--pressure (TP) structure of a generic ultra-hot Jupiter atmosphere (Table~\ref{table:parameters}) with the \texttt{HELIOS} radiative-convective equilibrium code \citep{malik2017helios, malik2019self}. We perform our calculation at solar metallicity and carbon-to-oxygen ratio (C/O), including gaseous metals and H$^-$ opacity, at a zenith angle of 83$\degr$. This angle is loosely where we expect the actinic flux (i.e., the flux capable of contributing to photochemistry) to reach an optical depth of 1 \citep{tsai2023photochemically}. The result of this calculation is shown in Figure~\ref{fig:helios_models}. The TP profile is convective at depth and exhibits a temperature inversion, typical for ultra-hot Jupiters \citep[e.g.,][]{haynes2015spectroscopic,evans2017ultrahot,sheppard2017evidence,lothringer2018extremely,coulombe2023broadband}. In our model, the temperature inversion is predominantly driven by a combination of TiO, VO (deeper than 1~mbar), Fe I (shallower than 1~mbar), Fe II (near 1~microbar), and alkalis (with contribution toward the top of the atmosphere\footnote{This result is similar to \cite{molliere2015model}, who found that alkalis can cause inversions in hot Jupiter atmospheres without TiO and VO when the C/O ratio is roughly 1. In their case, the alkalis produce inversions because the main molecular cooling agents are depleted through carbon-rich chemistry. In our hotter atmosphere, however, alkalis contribute to inversions at solar C/O due to depletion of these cooling agents via (thermal) dissociation.}).

\begin{figure}
    \centering
    \includegraphics[scale=0.55]{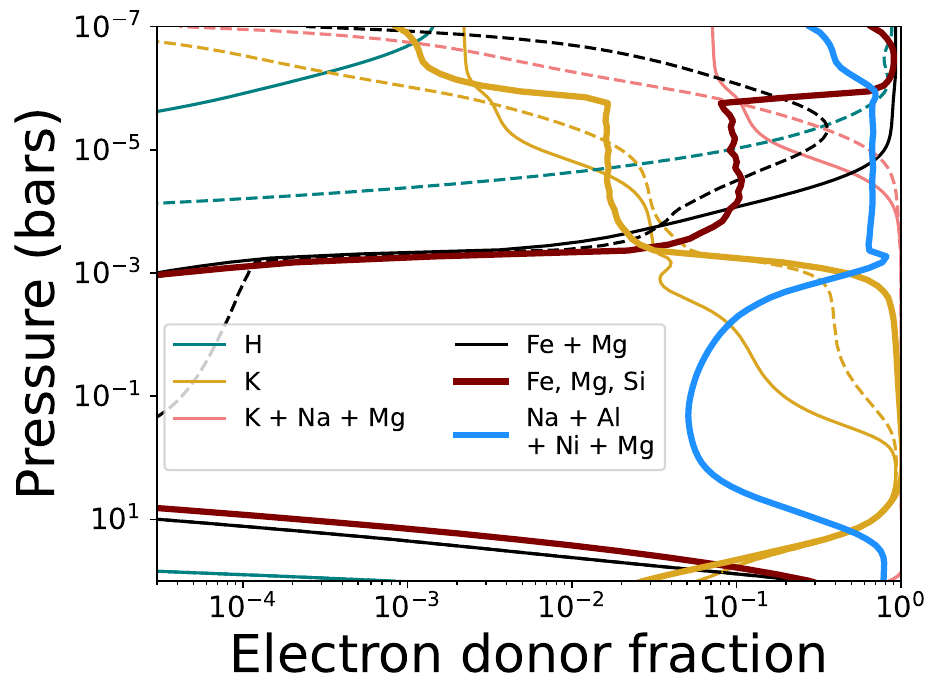}
    \caption{The primary electron donors in our chemical calculations. The thick solid lines represent equilibrium chemistry. Thin solid lines represent \texttt{VULCAN} disequilibrium chemistry with the F5V stellar spectrum, and thin dotted lines represent represent disequilibrium chemistry with the HD 189733 stellar spectrum.}
    \label{fig:donors}
\end{figure}

We next consider equilibrium chemistry by interpolating \texttt{FastChem} \citep{kitzmann2024fastchem} tables along our model TP profile, using fits to equilibrium constants including ions up to Uranium. This step produces the free electron, neutral gas, and Ba II number densities as a function of pressure. Under equilibrium chemistry, K is the primary electron donor between roughly 10 bars and 1 mbar (Figure~\ref{fig:donors}). Between $\sim$1 mbar and 1 microbar, Na, Al, Ni, and Mg together yield 2/3 of the total free electron population. Finally, below this pressure, the bulk of the electron donation is from Fe, Si, and Mg. The quantities calculated in this step, along with a given B field strength, allow us to calculate $u_i - u_n$ from Equation~\ref{eq:central_equation}.

Our model predicts that Ba II--neutral velocity differences well exceed 1 km\,s$^{-1}$ for reasonable B field values (Figure~\ref{fig:helios_models}).\footnote{Interestingly, our model also predicts that ions and neutrals are very well coupled deep in the atmosphere. This result essentially affirms the magnetization regimes laid out in \cite{koskinen2014electrodynamics}.} This method is evidently sensitive to B field strengths to a measurable degree.

We note that this method requires relatively strong constraints on the pressure that both the ion and neutral spectral lines probe. Uncertainties in the contribution function by a dex in pressure would correspond to uncertainties in the estimated B field of more than an order of magnitude. Even so, an order of magnitude constraint on an exoplanet B field would still be highly useful. The calculation also assumes that both the ion and neutral probe the same pressure level. This approximation holds so long as the measured velocity of the neutral corresponds to the neutral velocity at the measured \textit{pressure} of the ion, or vice versa. For instance, the calculation is valid if Ba II is measured at 1 microbar and Fe I at 10 microbars, and the Fe I flow is approximately constant with pressure below 10 microbars. Simulations would be able to assess how reasonable this scenario might be.

\begin{figure}
    \centering
    \includegraphics[scale=0.55]{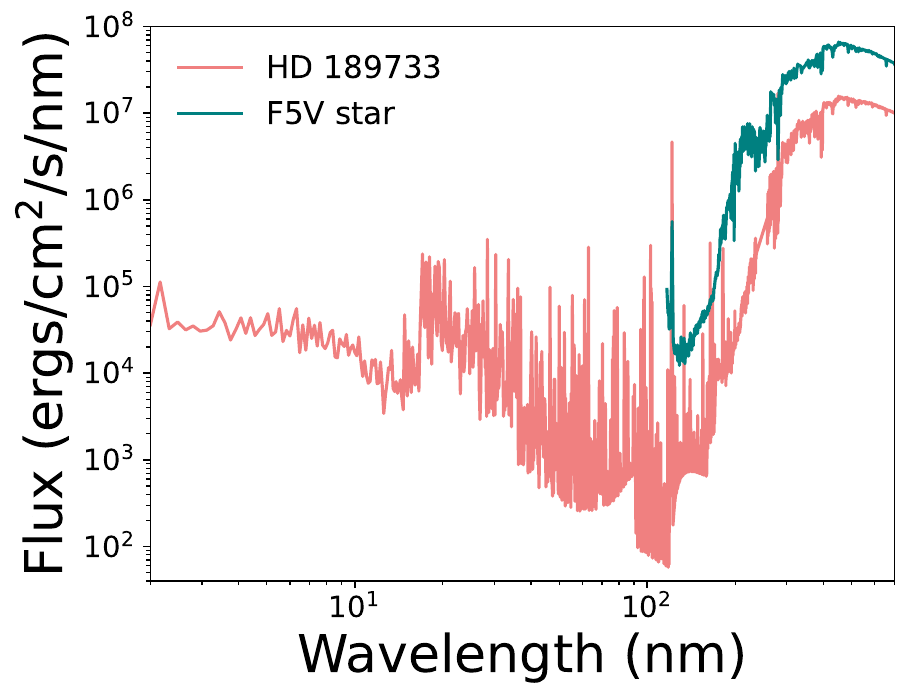}
    \caption{The stellar UV spectra input to \texttt{VULCAN}.}
    \label{fig:uv_spectra}
\end{figure}

Thus far, we have only considered thermal ionization in our model atmosphere. Highly irradiated gas giants may additionally be subject to photoionization at low pressures, which pushes the free electron number density out of thermal equilibrium \citep[e.g.,][]{yelle2004aeronomy,cho2008atmospheric,koskinen2010ionization, helling2021understanding}. To assess the impact of this uncertainty, we run a grid of \texttt{VULCAN} \citep{tsai2021comparative} photochemical simulations based on our \texttt{HELIOS} temperature structure. We use a chemical reaction network that includes neutrals \textit{and} ions for this calculation, as previously shown in \cite{lee2020simplified} and \cite{coulombe2023broadband}.\footnote{The official release of \texttt{ion-vulcan} is in preparation.} Varying the unknown eddy diffusion coefficient, $K_{\rm zz}$, by orders of magnitude, in addition to toggling molecular diffusion and changing the input stellar spectra (Table~\ref{table:parameters}, Figure~\ref{fig:uv_spectra}), consistently points to a single result: uncertainty in the XUV spectrum is the primary driver of uncertainty in our calculation. The strong scaling of $u_i - u_n$ with the planetary B field still allows order of magnitude estimates of the latter from the former, but quantitative estimates of the planetary B field will require estimates of the stellar flux at wavelengths shorter than $\sim 100$~nm.\footnote{Our chemical network does not currently include the ionization of neutral barium \citep[e.g.,][]{griesmann1992total} or of Ba II. Therefore, we cannot assess the effect of photoionization on the Ba II population itself. However, the intrinsic number density of this ion is so low that the contribution of Ba II to the total gas number density (the only way the abundance of Ba II features in Equation~\ref{eq:central_equation}) is very small. The $n_i$ assumed here under equilibrium chemistry could increase by 6 orders of magnitude before affecting $u_i - u_n$ at the percent level---making our results very robust to this source of uncertainty.} %Furthermore, Ba II itself could be photoionized into Ba III without affecting the calculating through $n_i$ itself.

When including photochemistry, K is still the main ion donor at $\sim$1 bar, but a combination of K, Na, and Ca dominates ion donation between 1 bar and almost 10~microbars. Below this pressure, H II is the major ion donor. This upper atmosphere behavior is the largest discrepancy with respect to equilibrium chemistry ion donation. Presumably, this effect arises because H$_2$ is photodissociated into atomic H \citep[e.g.,][]{yelle2004aeronomy, munoz2007physical}, which is then readily (photo)ionized. Note that this departure only occurs for the photochemical model with more XUV flux. In the other model, we revert to the equilibrium expectation that Mg and Fe predominantly donate ions in the upper atmosphere. Just as with the free electron population, the more XUV flux is considered, the greater the departure from equilibrium chemistry expectations. Broadly, photochemistry does not appear to have a very significant impact on our results. This robustness is likely due to the high temperatures of our ultra-hot Jupiter causing fast chemical reaction rates and a tendency toward chemical equilibrium.

To round out our proof of concept, we calculate the transmission spectrum of this atmosphere under chemical equilibrium. We do so with the transmission version of the 1D \texttt{CHIMERA} code \citep[e.g.,][]{line2013systematic}. We calculate our spectrum at a resolution of R = 300,000 and subsequently convolving it to R = 145,000, similar to the resolution of the ESPRESSO spectrograph \citep{pepe2021espresso}. We include Ba II, Fe, Mg, Ca II, Na, K, and VO as opacity sources.\footnote{All species' opacity data are drawn from \cite{kurucz2011including}. The exception is VO, which is drawn from \cite{mckemmish2016exomol}.} We also include Rayleigh scattering from H and He. We calculate the vertically varying equilibrium abundances of these gases according to the \texttt{HELIOS}-provided TP profile. We do not explicitly include H$^{-}$. H$^{-}$ opacity is roughly constant over this wavelength range, so we simply model it as a continuum opacity \citep[e.g.,][]{pelletier2023vanadium}. We set this continuum rather arbitrarily at 10 mbar, emulating the effect of a large free electron population.

\begin{figure*}
    \centering
    \includegraphics[scale=0.651]{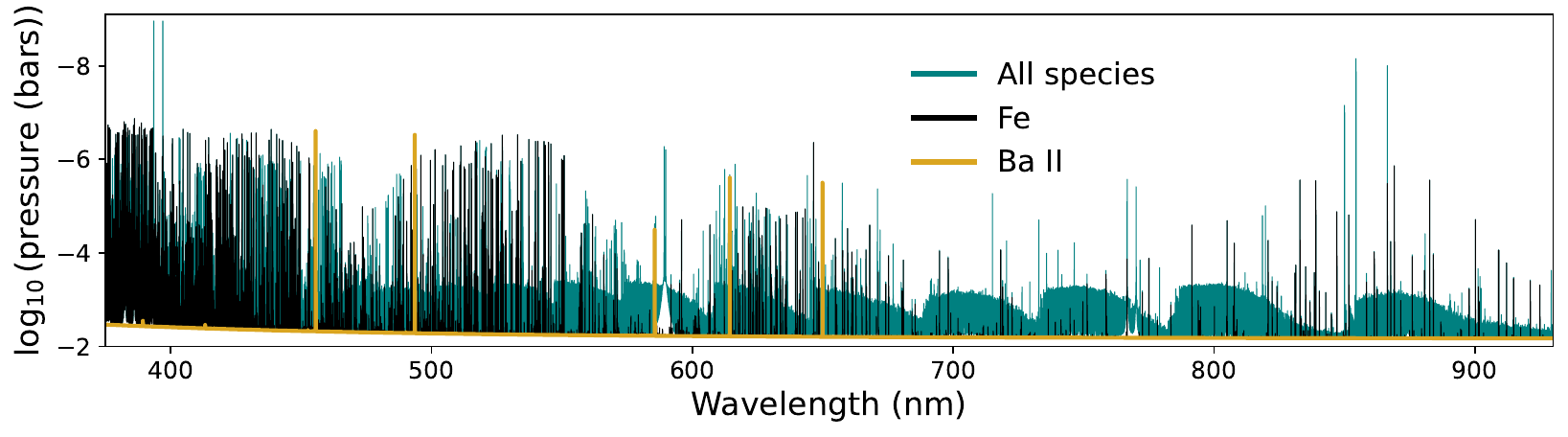}
    \caption{An R = 145,000 \texttt{CHIMERA} optical transmission spectrum of an ultra-hot Jupiter under chemical equilibrium. A spectrum with Ba II as the only spectrally active trace species is overplotted (gold), indicating the location and strength of Ba II absorption lines in our model atmosphere. Ba II lines in this model probe pressures near a microbar. We generate an analogous spectrum for neutral iron, as well (black).}
    \label{fig:spectrum}
\end{figure*}

We show the resultant spectrum in Figure~\ref{fig:spectrum}. It is clear that multiple Ba II lines peek out above the continuum, with a few of the strongest reaching a slant optical depth of unity at pressures less than 1 microbar. Referring back to Figure~\ref{fig:helios_models}, slip velocities evidently are expected to be measurable for a wide range of B field strengths within this pressure range. Fe appears to be a suitable neutral against which to compare; it has a number of strong lines that probe similar pressure levels.

\section{Conclusion} \label{sec:conclusion}
Magnetic fields in exoplanets are largely unconstrained by both theory and observation. Motivated by the need for more leverage on this fundamental quantity, we demonstrate that tracking heavy ion--neutral velocity differences is useful for estimating planetary magnetic fields. Because ions are deflected by magnetic fields and neutrals are not, strong differences between the two populations' velocities imply strong magnetic fields. A recommended recipe for applying this technique is as follows:

\begin{enumerate}
    \item Constrain the XUV flux of the host star (such as extrapolating measurements from  Hubble COS observations; e.g., \citealt{france2016muscles}).
    \item Assess whether the XUV flux is strong enough to appreciably drag Ba II in escaping flow (e.g., Equation~\ref{eq:critical_flux}).
    \item Measure the velocities of a suitable heavy ion (such as Ba II) and a suitable heavy neutral (such as Fe) with high-resolution spectroscopy. Ideally, this step is performed in a retrieval context, thus constraining the thermal structure of the atmosphere along with other potential aliases and/or degeneracies \citep[e.g.,][]{madhusudhan2009temperature,brogi2019retrieving,gandhi2019hydra,gibson2020detection, pelletier2023vanadium}.
    \item Calculate the electron density at the pressure level probed by the two gas species, ideally accounting for photoionization.
    \item With the available information, estimate the B field strength (Equation~\ref{eq:central_equation}).
\end{enumerate}

From an observational perspective, this approach is ready to be applied immediately. For a substantial range of B field strengths, the predicted velocity differences here are large enough to observe (Table~\ref{table:barium_detections}). Specifically, ultra-hot Jupiters orbiting bright stars appear to be particularly amenable to this technique, as their atmospheres are highly irradiated enough to support a substantial steady-state ion population and large enough to appreciably occult their host stars---in addition to exhibiting relatively low photon noise. Even for these planets, discrepancies have been seen between high-resolution measurements made with different instruments (e.g., Table~\ref{table:barium_detections}) or with different epochs on the same planet, which may be astrophysical or may be revealing some key differences in data processing procedures. Such issues may be resolved with a larger sample size: broader observational campaigns and archival efforts to measure heavy ion--neutral velocity differences could build up a population of such detections, with ramifications for understanding hot Jupiter interior structure and dynamo theory in an interesting forcing regime \citep[large energy flux under slow rotation;  e.g.,][]{christensen2010dynamo}. In particular, it would be valuable to assess the inferred B field strength as a function of equilibrium temperature and rotation, as both of these planetary properties could impact dynamo field generation \citep[e.g.,][]{kao2016auroral,yadav2017estimating,shulyak2017strong, brain2024exoplanet}.

Our proposed method benefits from the availability of relevant observations, as measured slip velocity precision (a few tenths to 1 km/s; Table~\ref{table:barium_detections}) appears amenable to order-of-magnitude B field constraints (Figure~\ref{fig:helios_models}).  Even so, ours is a simple toy model, so it currently neglects a number of confounding factors. Further theoretical work is needed to enable quantitative estimates of (potentially very diverse) exoplanetary magnetic fields. Avenues for future work include accounting for:
 
\begin{itemize}
    \item neutral flow feedback processes and microphysical regulating effects, which would likely act against superthermal ion flow and require a reformulation of the induction equation assumed in this work \citep[e.g.,][]{norman1978kinetic,hillier2024ambipolar, hopkins24micro}
    \item more explicit treatment of multi-fluid MHD \citep[e.g.,][]{wardle1999conductivity,hopkins24micro}, as Ba II is not the only ion species---and, in fact, is not the primary electron donor 
    \item B field geometric complexity, such as departures from a dipole in the deep-seated field (Jupiter has substantial quadrupole and octopole B field contributions; e.g., \citealt{acuna1976main}) or induced magnetic fields \citep[e.g.,][]{rogers2014magnetohydrodynamic,rogers2017hottest,soriano2023magnetic}
    \item tensor conductivity \citep[e.g.,][]{koskinen2014electrodynamics}
    \item the prospect of using multiple heavy ion--neutral pairs to calibrate B field estimates for a single planet
    \item more physical photochemistry, such as ionic ambipolar diffusion and the photoionization of Ba
    \item whether XUV reconstructions are sufficient for reducing the uncertainty in the free electron population \citep[e.g.,][]{youngblood2017muscles,melbourne2020estimating,teal2022effects}
    \item  3D ion--neutral velocity differences, especially as projected onto high-resolution measurements of line-of-sight velocities \citep[e.g.,][]{kempton2012constraining, showman2012doppler, flowers2019high, beltz2020significant, harada2021signatures,malsky2021modeling,wardenier2021decomposing,beltz2023magnetic,savel2023diagnosing,wardenier2023modelling}. At the poles, dipolar field lines are parallel to lines of sight, whereas they are perpendicular at the equator. It is not obvious how the latitude dependence of our effect averages over, e.g., the terminator
\end{itemize}

Of these considerations, one of the most crucial to address is the neutral flow feedback processes.  Our model is clearly not yet useful in the strong B field limit, where ion velocity must be limited by interactions with the bulk flow, because the ion velocities otherwise become unreasonably large.

Of similar importance are 3D effects. On one hand, 3D considerations could open up further avenues for investigation, with the potential for inferring not only exoplanetary B field strength, but also their \textit{orientation}, via phase-resolved ion--neutral velocity differences. On the other hand, ``3D-ness'' introduces complexities: velocity differences between gases can be caused by a variety of spatial variations \citep[e.g.,][]{ehrenreich2020nightside,wardenier2021decomposing,savel2022no,beltz2023magnetic,savel2023diagnosing, wardenier2023modelling}. This concern further motivates including self-consistent 3D forward models in high-resolution retrievals. If a 3D model can ``model out'' the slip velocity contribution due to, e.g., presence of Ba II on one limb and not the other, any residual slip velocity could be attributed to magnetic effects. This obstacle additionally presents an opportunity for using complementary methods for constraining magnetic field strength (e.g., measuring neutral flow velocity; e.g., \citealt{beltz2021exploring}). Agreement across multiple different B field measurement techniques would greatly increase confidence in a given constraint.

Looking forward: this technique is in principle applicable to any gas giant that is hot enough for sufficient ionization and is amenable to characterization with high-resolution spectroscopy. Especially once the extremely large telescopes (ELTs) are on-sky \citep[e.g.,][]{gilmozzi2007european,johns2012giant}, observers will be able to directly resolve a large number of exoplanetary spectral lines \citep[e.g.,][]{palle2023ground}, providing greater sensitivity to individual species' velocity profiles through line profiles \citep[e.g.,][]{seidel2021into}. These telescopes will additionally be able to target fainter host stars with high-resolution spectroscopy, thus expanding our planetary population even further. Moving beyond individual planets will potentially allow us to marginalize over unknown physics. Successfully doing so would reveal, at a basic level, how magnetic fields are generated in planetary bodies.

\begin{acknowledgments}
We thank Emma Mirizio for an informative conversation on aurorae. We also thank Madeline Lessard for benchmarking our calculations following \cite{thorngren2019intrinsic}. Additionally, we gratefully acknowledge Stefan Pelletier for providing cross-correlation maps for the planet WASP-76b as observed with MAROON-X. We thank Riley McDanal for helpful comments. Finally, we thank the reviewer for their feedback, which greatly improved thxe quality of this manuscript.
\end{acknowledgments}

\software{\texttt{astropy} (\citealt{astropy:2018}), \texttt{CHIMERA} \citep{line2013systematic}, \texttt{FastChem} \citep{kitzmann2024fastchem}, GitHub Copilot \citep{chen2021evaluating}, \texttt{HELIOS} \citep{malik2017helios}, \texttt{HELIOS-K} \citep{grimm2021helios},  \texttt{IPython} (\citealt{perez2007ipython}), \texttt{Matplotlib} (\citealt{hunter2007matplotlib}),  \texttt{Numba} \citep{lam2015numba}, \texttt{NumPy} (\citealt{2020NumPy-Array}),
\texttt{pandas} (\citealt{mckinney2010data}), 
\texttt{SciPy} (\citealt{virtanen2020scipy}), \texttt{tqdm} (\citealt{da2019tqdm}), \texttt{VULCAN} \citep{tsai2021comparative}}

\bibliography{sample631}{}
\bibliographystyle{aasjournal}



\end{document}